 \documentclass[twoside,reqno,11pt]{fcaa}

\usepackage{graphicx}
\usepackage{epsfig}
\usepackage{amsthm}
\usepackage{amsmath}
\usepackage{latexsym}
\usepackage{amsfonts}
\usepackage{amssymb}

\newtheoremstyle{theorem}
  {15pt}          
  {15pt}  
  {\sl}  
  {\parindent}
  {\sc}  
  {. }    
  { }    
  {}     
\theoremstyle{theorem}

\newtheoremstyle{defi}
  {15pt}          
  {15pt}  
  {\rm}  
  {\parindent}     
  {\sc}  
  {. }    
  { }    
  {}     
\theoremstyle{defi}

 \def\proofname{\indent\rm P\ r\ o\ o\ f.}
 
 \def\qed{\hfill$\Box$} 

 \usepackage{float}
 \usepackage{hyperref} 

    \def\themycopyrightfootnote{
       \hspace*{-0.7cm}
   }

 \title[Symmetric Erd$\acute{\text{E}}$lyi-Kober type operators]
 {Reflection symmetric Erd$\acute{\text{E}}$lyi-Kober type operators - \\ [2pt]
  A quasi-particle interpretation}
 \author[R. Herrmann]{Richard Herrmann}

\begin{document}

 \vbox to 2.5cm { \vfill }


 \bigskip \medskip

 \begin{abstract}
Reflection symmetric Erd$\acute{\text{e}}$lyi-Kober type fractional integral operators are used to construct fractional quasi-particle generators. The eigenfunctions and eigenvalues of these operators are given analytically. 
A set of fractional creation- and annihilation-operators is defined and the properties of the corresponding  free Hamiltonian are investigated. Analogue to the classical approach for  interacting multi-particle systems the results are  interpreted as a fractional quantum model  for a description of residual interactions of pairing type. 
 \ 
 \medskip

 \smallskip


 \end{abstract}

 \maketitle

 \vspace*{-16pt}



\section{Introduction}
The physical interpretation of operators used within the framework of  fractional calculus is a vividly discussed area of actual research \cite{kir94,pod03,her14b}. One complication is the fact, that different integral- and differential operators are equally reasonable and different levels of complexity are involved, like e.g. Riesz- \cite{rie49, fel52} and Erd\'elyi-Kober \cite{erd40,kob40,sne66, pag12} integrals respectively. 
   
The  Riesz integral ${_\textrm{\tiny{RZ}}}I^\alpha$  is given as the convolution integral, depending on a single fractional parameter $\alpha$:
\begin{eqnarray}
{_\textrm{\tiny{RZ}}}I^\alpha f(x) \ 
&=&
\label{RZ}
\frac{1}{2 \sin(\pi\alpha/2 ) \Gamma(1 - \alpha)}
     \int_{-\infty}^{\infty} \!\!\!  d\xi \, d^{-\alpha}
     f(\xi)\\
&=&
\frac{1}{2 \sin(\pi\alpha/2 ) \Gamma(1 - \alpha)}
     \int_{-\infty}^{\infty} \!\!\!  d\xi \, |x-\xi|^{-\alpha}
     f(\xi), \, 0 < \alpha < 1
\end{eqnarray}
with a weakly singular kernel of power-law type, where $d = | x - \xi |$ is a measure of distance on $R^1$. 

In the following we will discuss a specific generalization of this integral, based on  a Cassini-type kernel, which is the product of two measures of distance and is considered as a symmetric extension of the one-dimensional Erd\'elyi-Kober type integral. A family of generalized fractional integrals of this type has been introduced recently \cite{her14b}. 

In this paper, we will derive the eigenvalues and eigenfunctions of this operator.   
We will collect arguments in support of the idea, that there is a close relationship between the proposed generalized fractional operator and  the concept of quasi-particle operators, widely used in solid states physics, in the theory of super-conductivity and in nuclear physics respectively to model pairing effects.   

For that purpose, we will first introduce the symmetric one-dimensional Erd\'elyi-Kober type integral and give an analytic derivation of its main properties.  We will present reasonable definitions of corresponding creation- and annihilation-operators and finally investigate the properties of the free Hamiltonian based on these generalized fractional operators and interpret its properties in close analogy to Cooper-pairs in momentum representation.

\section{Eigenfunctions and eigenvalues of the reflection symmetric one-dimensional Erd\'elyi-Kober type integral}
We will investigate a special case of the one-dimensional Erd\'elyi-Kober type integral, which contains two foci of the form 
\begin{eqnarray}
\label{ek00}
{_\textrm{\tiny{EK}}}I^{\alpha,\gamma}  f(x) &=&
{_\textrm{\tiny{EK}}}\mathcal{N}
     \int_{-\infty}^{\infty} \!\!\!  d\xi \, d_1^{-\alpha/2} \,  d_2^{-\gamma/2}  f(\xi), 
\qquad 0 < \alpha + \gamma < 2 
\end{eqnarray}
where  
\begin{equation}
d_i = |x_i-\xi| \qquad\qquad\qquad\qquad i=1, 2
\end{equation}
determines the distance between a focal point $x_i$ and a position $\xi$,  $\mid\cdot\mid$ denotes the absolute value and the normalization factor 
${_\textrm{\tiny{EK}}}\mathcal{N}$ is chosen such 
\begin{equation}
{_\textrm{\tiny{EK}}}\mathcal{N}
=
\frac{1}{2 \sin(\pi(\alpha/4 + \gamma/4) ) \Gamma(1 - \alpha/2 - \gamma/2)} 
\end{equation}
that in the case $\gamma \rightarrow 0$ up to a scaling in $\alpha$ the integral coincides with the Riesz integral definition.  

Setting $\gamma = \alpha$  and defining the focal points $x_i$ by introducing their distance $p$  via the reflection symmetric relation with respect to  $x$
\begin{equation}
x_1 = x -  \frac{p}{2} \qquad x_2 = x + \frac{p}{2}  
\end{equation}
we obtain a reflection symmetric one-dimensional Erd\'elyi-Kober type integral operator with Cassini kernel type:
\begin{eqnarray}
\label{sekp}
{_\textrm{\tiny{EK}}}I^{\alpha, p}  f(x) &=&
\frac{1}{2 \sin(\pi\alpha/2 ) \Gamma(1 - \alpha)}  \times  \nonumber \\
& & 
     \int_{-\infty}^{\infty} \!\!\!  d\xi \,
\big(
|x-\frac{p}{2}-\xi|^{-\alpha/2} \,|x+\frac{p}{2}-\xi|^{-\alpha/2}
\big)
     f(\xi) \\
& &  \qquad \qquad \qquad\qquad \qquad\qquad   \,\, p \in \mathbb{R}, \, 0< \alpha<1 \nonumber 
\end{eqnarray}
which in the limit $p \rightarrow 0$ reduces to the Riesz fractional integral.
  
The eigenfunctions of this operator are determined by the integral equation 
\begin{eqnarray}
\label{eigen11}
{_\textrm{\tiny{EK}}}I^{\alpha, p}  \psi(x) &=&
\Theta(k,p) \psi(x)
\end{eqnarray}
and are both given as trigonometric functions $\psi(x) = \{ \cos(k x), \sin(k x)  \}, k \in \mathbb{R}$. The corresponding  eigenvalues $\Theta(k,p)$ result with the help of (4.3.17, 9.1.20, 9.1.24)  from \cite{abr65} in
\begin{eqnarray}
\label{eigen12}
\Theta(k,p) &=&
\frac{ \sqrt \pi \, 
 \Gamma(1-\alpha/2)}{2 \sin(\pi\alpha/2 ) \Gamma(1 - \alpha)}  \times  |k/p|^{(\alpha-1)/2} \times \\
& & \qquad
     \big(J_{(1-\alpha)/2}(|k p|/2) -  Y_{(\alpha-1)/2}(|k p|/2)\big) \nonumber\\
& & \qquad\qquad\qquad\qquad\qquad\qquad\qquad k, p \in \mathbb{R}, 0< \alpha<1\nonumber
\end{eqnarray}
where $J_\beta(z)$ and $Y_\beta(z)$ denote the corresponding standard Bessel functions.

\noindent
The following asymptotic cases clarify the functional behavior of the eigenvalue spectrum $\Theta(k,p)$:
 \begin{itemize}
 \item  Limit for $p \rightarrow 0$

A series expansion of the eigenvalue spectrum for small $p$ yields:
\begin{eqnarray}
\label{eigen14}
\Theta(k,p) &\approx&
     |k|^{\alpha-1} \! \! -
     |p|^{1-\alpha}\frac
{4^\alpha \pi \sin^2(\pi \alpha/4)  }
{\sin(\pi \alpha) \Gamma((1-\alpha)/2)\Gamma((3-\alpha)/2)}  \nonumber \\
& & 
+ o(|p|^{2-\alpha}) 
\qquad k, p \in \mathbb{R}, \, |p| \ll 1, \,  0< \alpha<1
\end{eqnarray}
which is well behaved for small $p$ within the range of allowed $\alpha$-values and therefore  a smooth transition to the eigenvalue spectrum of the Riesz integral results for vanishing $p$:
\begin{eqnarray}
\label{eigen15}
\lim_{p \rightarrow 0} \Theta(k,p) &=&
     |k|^{\alpha-1} 
 \qquad\qquad k, p \in \mathbb{R},  \, 0< \alpha<1
\end{eqnarray}
\item  Limit for $|k p|  \rightarrow \infty$

The asymptotic expansion of the eigenvalue spectrum for $|kp| \gg 1$, which includes the cases $k \gg 1$ and $p $ fixed and $p \gg 1$ and $k$ fixed,   results according to (9.2.1, 9.2.2) from \cite{abr65} in
\begin{eqnarray}
\label{eigen16}
\Theta^\infty(k,p) &\approx&
\frac{\Gamma(1-\alpha/2)}
{\sin(\pi\alpha/4 ) \Gamma(1 - \alpha)}
|k|^{\alpha/2-1}|p|^{-\alpha/2} \cos(k p/2) \nonumber \\
& & + o(|k p|^{-1}) \qquad\quad k,  p \in \mathbb{R},\, |k p| \gg 1,\,  0< \alpha<1
\end{eqnarray} 
which determines a damped oscillatory behavior for increasing $|k p|$.

\item Special case $p = 1/k$

A remarkable property of the eigenvalue spectrum  results from the periodicity of the trigonometric functions. For the special case $p = 1/k$ the eigenvalue spectrum  is given by 
\begin{eqnarray}
\label{eigen15}
\Theta(k,1/k) &=&
\frac{\sqrt \pi \, 
 \Gamma(1-\alpha/2)}{2 \sin(\pi\alpha/2 ) \Gamma(1 - \alpha)}    \times      |k|^{(\alpha-1)}
\times \\
& & \qquad
     \big(J_{(1-\alpha)/2}(1/2) -  Y_{(\alpha-1)/2}(1/2)\big) \nonumber \\
& = & 
     c(\alpha) |k|^{(\alpha-1)}
\qquad\qquad\qquad k \in \mathbb{R}, \, 0< \alpha<1
\end{eqnarray} 
and therefore coincides up to a factor $c(\alpha) $ with the Riesz fractional integral eigenvalue spectrum determined for trigonometric functions. Consequently the use of different kernel functions may lead to similar spectra. 

\end{itemize}
\begin{figure}
\begin{center}
\includegraphics[width=\textwidth]{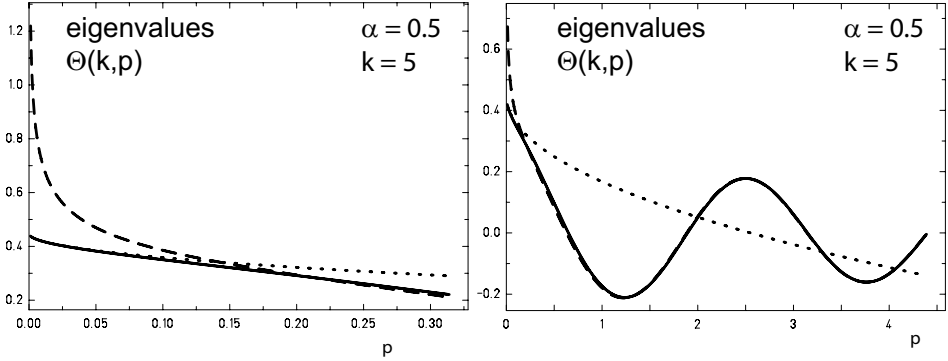}
\caption{Eigenvalue as a function of the distance $p$. On the left  for small $p$, on the right for larger $p$. Thick line indicates the exact solution (\ref{eigen12}), pointed line the low-distance approximation (\ref{eigen14}) and dashed line the asymptotic expansion (\ref{eigen16}).}
\label{fig1}
\end{center}
\end{figure}

In figure \ref{fig1} we have plotted the eigenvalue spectrum and the corresponding approximations. In case $k p > 1$ the asymptotic expansion (\ref{eigen16}) is a good approximation to the exact solution, which shows an oscillatory behavior as a function of $k$ and $p$ respectively. 

For a direct explanation, we consider this behavior dominated by  the result of an  interference of two contributions, since      
with 

\begin{equation}
\label{eigen23}
\cos\bigl(k(x- \frac{p}{2})\bigr)+ \cos\bigl(k(x+ \frac{p}{2})\bigr) = 2 \cos(k  p/2)\cos(k x) 
\end{equation} 
the asymptotic limit (\ref{eigen16}) may be rewritten as  
\begin{eqnarray}
\label{eigen24}
\qquad \Theta^\infty(k,p) &&\!\!\!\!\!\!\!\!\!\!  \cos(k x) = \nonumber \\
&&
 \frac{1}{2} \Bigl(
\frac{\Gamma(1-\alpha/2)}
{\sin(\pi\alpha/4 ) \Gamma(1 - \alpha)}
|k|^{\alpha/2-1}|p|^{-\alpha/2} 
 \cos\bigl(k(x- \frac{p}{2})\bigr) 
\nonumber \\
&& + 
\frac{\Gamma(1-\alpha/2)}
{\sin(\pi\alpha/4 ) \Gamma(1 - \alpha)}
|k|^{\alpha/2-1}|p|^{-\alpha/2} 
 \cos\bigl(k(x+ \frac{p}{2})\bigr)\Bigr) \nonumber \\
& & + o(|k p|^{-1}) \qquad\quad k,  p \in \mathbb{R},\, |k p| \gg 1,\,  0< \alpha<1
\end{eqnarray} 
which may be interpreted as the result of a superposition of two shifted single focus kernel integrals of Riesz type. 

In close analogy to Debye's saddle-point approximation for standard integrals \cite{deb09}  this can be derived in full generality  from the symmetric Erd$\acute{\text{e}}$lyi-Kober kernel via a Taylor series expansion of $d_1$ at $x_2$ and and of $d_2$ at $x_1$:
\begin{eqnarray}
\label{eigen25}
d_1^{-\alpha/2}|_{x_2} = |x_1-\xi|^{-\alpha/2}|_{x_2} &\approx&  |p|^{-\alpha/2}   - o\bigl(|p|^{-2-\alpha/2} (\xi-x_2)\bigr)\\
d_2^{-\alpha/2}|_{x_1} = |x_2-\xi|^{-\alpha/2}|_{x_1} &\approx&  |p|^{-\alpha/2}  + o\bigl(|p|^{-2-\alpha/2} (\xi-x_1)\bigr) \\
& &\qquad \qquad p \in \mathbb{R},\, |p| \gg 1,\,  0< \alpha<1 \nonumber
\end{eqnarray} 
and therefore
\begin{eqnarray}
\label{eigen36}
{_\textrm{\tiny{EK}}}I^{\alpha, p}  f(x) &=&
{_\textrm{\tiny{EK}}}\mathcal{N}
   \int_{-\infty}^{\infty} \!\!\!  d\xi \,
d_1^{-\alpha/2} \,d_2^{-\alpha/2}      f(\xi) \\
 &\approx&
{_\textrm{\tiny{EK}}}\mathcal{N}
     \int_{-\infty}^{\infty} \!\!\!  d\xi \,
(d_1^{-\alpha/2} |p|^{-\alpha/2} + d_2^{-\alpha/2} |p|^{-\alpha/2})    f(\xi) \\
 &=&
{_\textrm{\tiny{EK}}}\mathcal{N}
|p|^{-\alpha/2} \bigl(
     \int_{-\infty}^{\infty} \!\!\!  d\xi \,
d_1^{-\alpha/2}    f(\xi)
+
     \int_{-\infty}^{\infty} \!\!\!  d\xi \,
 d_2^{-\alpha/2}     f(\xi) \bigr) \\
 &=&|p|^{-\alpha/2} {_\textrm{\tiny{EK}}}\mathcal{N}
     \int_{-\infty}^{\infty} \!\!\!  d\xi \,
|x-\xi|^{-\alpha/2}    f(\xi+p/2)
\nonumber \\
& &
+|p|^{-\alpha/2}{_\textrm{\tiny{EK}}}\mathcal{N} 
     \int_{-\infty}^{\infty} \!\!\!  d\xi \,
 |x-\xi|^{-\alpha/2}     f(\xi-p/2) \\
& &  \qquad \qquad \qquad\qquad \qquad\qquad   \,\, p \in \mathbb{R}, \, |p| \gg 1,  \, 0< \alpha<1 \nonumber 
\end{eqnarray}
which is up to a factor the sum of two shifted Riesz integrals.

This observation directly leads to a physical interpretation of the  symmetric Erd$\acute{\text{e}}$lyi-Kober integral operator in terms of  a two-particle operator.

\section{Creation- and annihilation-operators of symmetric Erd$\acute{\text{e}}$lyi-Kober type}
Let us define the non-local pendant ${_\textrm{\tiny{C}}}\hat{O}$ of Caputo type of a local space-dependent operator ${_\textrm{\tiny{loc}}}\hat{o}$ based on the symmetric Erd$\acute{\text{e}}$lyi-Kober type integral (\ref{ek00})  as 
\begin{eqnarray}
\label{EKCC}
{_\textrm{\tiny{C}}}\hat{O} &=&
{_\textrm{\tiny{EK}}}I^{\alpha,\gamma}  
{_\textrm{\tiny{loc}}}\hat{o}
\end{eqnarray}
and of Riemann type with inverted operator sequence
\begin{eqnarray}
\label{EKRR}
{_\textrm{\tiny{R}}}\hat{O} &=&
{_\textrm{\tiny{loc}}}\hat{o}\,
{_\textrm{\tiny{EK}}}I^{\alpha,\gamma}  
\end{eqnarray}

The quantization of a multi-particle system may be realized on a Fock-space introducing creation- ($\hat{c}^\dagger$) and  annihilation- ($\hat{c}$) operators acting on a vacuum state  $\mid0>$, which is characterized by the condition:
\begin{eqnarray}
\label{EKvacuum}
\hat{c} \mid 0 >  &=& 0
\end{eqnarray}
In occupation number representation the Fock-space is constructed by repeated action of the creation-operator on a given state $\mid n > $
\begin{eqnarray}
\hat{c}^\dagger \mid n >  &=& \sqrt{n + 1}\mid n + 1> \\
\hat{c} \mid n >  &=& \sqrt{n}\mid n -1> 
\end{eqnarray}
We may therefore introduce the following fractional creation and annihilation-operators:
 
According to (\ref{EKCC}) we define a set of Caputo type operators in space representation
\begin{eqnarray}
\label{EKC11}
{_\textrm{\tiny{C}}}\hat{C}^\dagger   &=& {_\textrm{\tiny{EK}}}I^{\alpha,\gamma}  \hat{c}^\dagger  \\
{_\textrm{\tiny{C}}}\hat{C}   &=& {_\textrm{\tiny{EK}}}I^{\alpha,\gamma}  \hat{c} 
\end{eqnarray}
and we introduce a set of Riemann type operators in space representation
according to (\ref{EKRR})
\begin{eqnarray}
\label{EKR11}
{_\textrm{\tiny{R}}}\hat{C}^\dagger   &=&  \hat{c}^\dagger  \, {_\textrm{\tiny{EK}}}I^{\alpha,\gamma}  \\
{_\textrm{\tiny{R}}}\hat{C}   &=&  \hat{c} \, {_\textrm{\tiny{EK}}}I^{\alpha,\gamma}  
\end{eqnarray}
The corresponding  non-local vacuum state  is determined by the condition:
\begin{eqnarray}
\label{EKRCV}
{_\textrm{\tiny{C}}}\hat{C}   \mid 0>{_\textrm{\tiny{C}}} &=&  0\\
{_\textrm{\tiny{R}}}\hat{C}   \mid 0>{_\textrm{\tiny{R}}}&=& 0
\end{eqnarray}
where $ \mid 0>{_\textrm{\tiny{C}}}$ and  $\mid 0>{_\textrm{\tiny{R}}}$ are the vacuum states associated with the Caputo- and Riemann-type operators, respectively.
 
It should be noted, that for the Caputo type set of operators (\ref{EKC11}) the vacuum state $\mid 0>{_\textrm{\tiny{C}}}$ is identical with its local pendant. For the Riemann type operators, in general the corresponding non-local vacuum state differs from its local pendant. 

\section{Quasi-particle interpretation}
In literature, classical applications of fractional calculus have been widely studied for time-like and space-like variables $x$, see e.g.  \cite{hil00, pod03,her14a}. 

The transition from classical mechanics to quantum mechanics may be interpreted as a transition from independent coordinate space and momentum space first to a Hilbert and then to a Fock space respectively, where space and momentum operators are treated similarly as a canonical coordinate set. 

Consequently a postulate of quantum mechanics states, that results must be independent of the specific choice of a representation e.g. space or momentum representation.  This is the mathematical manifestation of wave-particle-duality: The behavior of a given quantum object may be described completely using e.g. either the position $x$ or wave vector $k$ operator respectively.

As a direct consequence,  a fractional extension of standard quantum mechanics has to be formulated on a dual space. A treatment of time- and/or space dependent problems in fractional quantum mechanics using fractional operators e.g. Erd$\acute{\text{e}}$lyi-Kober type operators is  therefore based on the following representation pairs:    
\\
\begin{description}
\item[coordinate: time $t$, conjugate coordinate: energy $E$]
\begin{equation}
\{t,E\} =\{t,-i\hbar \partial_t \} \rightarrow  {_\textrm{\tiny{EK}}}I^{\alpha,\gamma}(t)  \{t, -i\hbar \partial_t \}
\end{equation}
For time-like coordinates the focal points $x_1 \equiv t_1$ and $x_2 \equiv t_2$ determine system properties earlier or later with respect to present $x \equiv t$, which touches the question of causal and anti-causal event sequences. 
The corresponding non-local phenomena are memory- or hysteresis-effects and are observed e.g. in magnetization curves of ferro-magnets \cite{mie03}. 

In order to obey the correct sequence of cause and effect  the symmetric Erd$\acute{\text{e}}$lyi-Kober type operator  may be used to describe particle / anti-particle pairs \cite{her14a}.

Since according to de Broglie the energy of a quantum object is directly related to a frequency via
\begin{equation}
E = \hbar \omega = h \nu
\end{equation}
\noindent
We may alternatively consider: 
\\

\item[coordinate: frequency $\nu$, conjugate coordinate: time $t$]
\begin{equation}
\{\nu,t\} =\{\nu,-ih \partial_\nu \} \rightarrow  {_\textrm{\tiny{EK}}}I^{\alpha,\gamma}(\nu)  \{\nu, -ih \partial_\nu \}
\end{equation}
In frequency representation, which is the canonically conjugate to time representation,  the points $x_1 \equiv \nu_1$ and $x_2 \equiv \nu_2$ mark the properties of lower or higher frequency with respect to $x \equiv \nu$.  Corresponding non-local concepts base on deviations of temporal coherence phenomena, which results in temporal interference or in the case of light, the deviation from monochromaticity.  

A key experiment is the Michelson-Morley-experiment, which failed to demonstrate the effect  on the speed of light relative to the ether. \cite{mic87}.   
\\

\item[coordinate: space $x$, conjugate coordinate: momentum $k$]
\begin{equation}
\{x,k\} =\{x,-i\hbar \partial_x \} \rightarrow  {_\textrm{\tiny{EK}}}I^{\alpha,\gamma}(x)  \{x,-i\hbar \partial_x \}
\end{equation}
For space-like coordinates the points $x_1, x_2$ mark a position  left or right with respect to $x$. The corresponding phenomena are non-local concepts also known historically as action-at-a-distance, which was suspected for centuries
by physicists and seemed obsolete after Maxwell introduced fields instead of forces, see also \cite{ein35}. 

In quantum mechanics it is known as quantum entanglement and marks the fact, that quantum numbers of quantum systems are correlated over large distances.  

A key-experiment on non-locality is e.g. the Aharonov-Bohm-effect \cite{aha59}. 
In \cite{her14b} we suggested, to use the symmetric Erd$\acute{\text{e}}$lyi-Kober type operator to describe the behavior of symmetric mesons, like strangeonium, charmonium and bottomoniom respectively. 

An alternative representation is:
\\

\item[coordinate: momentum $k$, conjugate coordinate: space $x$]
\begin{equation}
\{k,x\} =\{k,-i\hbar \partial_k \} \rightarrow  {_\textrm{\tiny{EK}}}I^{\alpha,\gamma}(k)  \{k,-i\hbar \partial_k \}
\end{equation}
In momentum representation, which is the canonically conjugate to coordinate representation,  the points $x_1 \equiv k_1$ and $x_2 \equiv k_2$ mark properties of slower or faster with respect to actual momentum $x \equiv k$.  Corresponding non-local concepts base on deviations of spatial coherence phenomena, which result in spatial interference. 

A key experiment is Young's interference/double slit experiment \cite{you02}.  

In case of angular momentum space, which may be considered the canonically conjugate to spherical coordinate space  e.g. in two dimensions the points $x_1 \equiv m_1$ and $x_2 \equiv m_2$ mark properties of slower or faster rotation with respect to actual angular momentum $x \equiv m$.    
\end{description}

In this section, we will give an example of a reasonable application of the  above introduced  Erd$\acute{\text{e}}$lyi-Kober type derivatives in terms of a momentum representation characterized by quantum numbers $k,p$. 

We will use the derived operators for a description of quantum multi-particle systems and apply
the concept of second quantization to model effects, which are observed e.g. in laser physics
as laser-electron interaction (excitons) \cite{kit87},  in low temperature physics as superconductivity (Cooper-pairs) \cite{coo56, bar57} and in nuclear physics as the pairing gap, which is observed as an odd-even effect in the binding energies of nucleons (pairing) \cite{bel59} or the back-bending effects for high-spin states in nuclei respectively \cite{gre96}, only to name a few.  

We will collect arguments which support the approach, that the properties of the hitherto presented complex fractional operators may be directly understood within the framework of such models.  

We will investigate the properties of  the diagonal Hamiltonian $H$
\begin{eqnarray}
\label{EHHH}
H = \sum_{k,p>0} E_{k,p}  \,  \hat{C}^\dagger_{p}(k) \, \hat{C}_{p}(k) 
\end{eqnarray}
where the non-local  creation- and annihilation-operators  of symmetric Erd$\acute{\text{e}}$lyi-Kober type are explicitly given e.g. for Caputo type operators  
\begin{eqnarray}
\label{eigen33c}
\hat{C}^\dagger_{p}(k) 
&=&
\frac{1}{2 \sin(\pi\alpha/2 ) \Gamma(1 - \alpha)}  \times  \nonumber \\
& &  
     \int_{-\infty}^{\infty} \!\!\!  d\kappa \,
\big(
|k-\frac{p}{2}-\kappa|^{-\alpha/2} \,|k+\frac{p}{2}-\kappa|^{-\alpha/2}
\big)\,      \hat{c}^\dagger_{\kappa} \\
\hat{C}_{p}(k) 
&=&
\frac{1}{2 \sin(\pi\alpha/2 ) \Gamma(1 - \alpha)}  \times  \nonumber \\
& & 
     \int_{-\infty}^{\infty} \!\!\!  d\kappa \,
\big(
|k-\frac{p}{2}-\kappa|^{-\alpha/2} \,|k+\frac{p}{2}-\kappa|^{-\alpha/2}
\big)\,      \hat{c}_{\kappa} \\
& & 
\qquad \qquad \qquad\qquad \qquad\qquad   \,\, p \in \mathbb{R}, \, 0< \alpha<1 \nonumber 
\end{eqnarray}
which we will use in the following derivation. Since the procedure for Riemann-type operators according (\ref{EKR11}) leads to similar results, it will be omitted here. 

For $p \gg 1$ we may replace the integral according to (\ref{eigen36}) by
\begin{eqnarray}
\label{eigen34}
\hat{C}^\dagger_{p}(k) 
&\approx&
\frac{1}{2 \sin(\pi\alpha/2 ) \Gamma(1 - \alpha)}  \times  \nonumber \\
& &  p^{-\alpha/2}
\int_{-\infty}^{\infty} \!\!\!  d\kappa \,
\bigl(
|k-\frac{p}{2}-\kappa|^{-\alpha/2}   + 
|k+\frac{p}{2}-\kappa|^{-\alpha/2} \bigr)\,      \hat{c}^\dagger_{\kappa}   \\
&=&
\frac{1}{2 \sin(\pi\alpha/2 ) \Gamma(1 - \alpha)}  \times  \nonumber \\
& &  p^{-\alpha/2}
\int_{-\infty}^{\infty} \!\!\!  d\kappa \,
\bigl(
|k-\kappa|^{-\alpha/2} \,      \hat{c}^\dagger_{\kappa-\frac{p}{2}}  + 
|k-\kappa|^{-\alpha/2} \,      \hat{c}^\dagger_{\kappa+\frac{p}{2}}  \bigr) \\
&=&
\frac{p^{-\alpha/2}}{2 \sin(\pi\alpha/2 ) \Gamma(1 - \alpha)} 
\int_{-\infty}^{\infty} \!\!\!  d\kappa \,
|k-\kappa|^{-\alpha/2}  \bigl(
\hat{c}^\dagger_{\kappa-\frac{p}{2}}  + 
 \hat{c}^\dagger_{\kappa+\frac{p}{2}}  \bigr) 
\end{eqnarray}
and similarly
\begin{eqnarray}
\label{eigen35}
\hat{C}_{p}(k) 
&\approx&
\frac{p^{-\alpha/2}}{2 \sin(\pi\alpha/2 ) \Gamma(1 - \alpha)} 
\int_{-\infty}^{\infty} \!\!\!  d\kappa \,
|k-\kappa|^{-\alpha/2}  \bigl(
\hat{c}_{\kappa-\frac{p}{2}}  + 
 \hat{c}_{\kappa+\frac{p}{2}}  \bigr) 
\end{eqnarray}
We now introduce four operators of shifted Riesz-Caputo-type in momentum space  via
\begin{eqnarray}
\label{eigen46}
\hat{A}^\dagger_{p}(k)
&=&
\frac{1}{2 \sin(\pi\alpha/4 ) \Gamma(1 - \alpha/2)}  
\int_{-\infty}^{\infty} \!\!\!  d\kappa \,
|k-\kappa|^{-\alpha/2} \,      \hat{c}^\dagger_{\kappa+\frac{p}{2}}   \\
\hat{A}_{p}(k) 
&=&
\frac{1}{2 \sin(\pi\alpha/4 ) \Gamma(1 - \alpha/2)}  
\int_{-\infty}^{\infty} \!\!\!  d\kappa \,
|k-\kappa|^{-\alpha/2} \,      \hat{c}_{\kappa+\frac{p}{2}}   \\
\hat{B}^\dagger_{p}(k) 
&=&
\frac{1}{2 \sin(\pi\alpha/4 ) \Gamma(1 - \alpha/2)}  
\int_{-\infty}^{\infty} \!\!\!  d\kappa \,
|k-\kappa|^{-\alpha/2} \,      \hat{c}^\dagger_{\kappa-\frac{p}{2}}   \\
\hat{B}_{p}(k) 
&=&
\frac{1}{2 \sin(\pi\alpha/4 ) \Gamma(1 - \alpha/2)}  
\int_{-\infty}^{\infty} \!\!\!  d\kappa \,
|k-\kappa|^{-\alpha/2} \,      \hat{c}_{\kappa-\frac{p}{2}}   \\
& & 
\qquad \qquad \qquad\qquad \qquad\qquad   \,\, p \in \mathbb{R}, \, 0< \alpha<2 \nonumber 
\end{eqnarray}
where
\begin{eqnarray}
\label{eigen56}
\hat{B}^\dagger_{p}(k)
&=&
\hat{A}^\dagger_{-p}(k) \\
\hat{B}_{p}(k)
&=&
\hat{A}_{-p}(k) \\
& & 
\qquad \qquad \qquad\qquad \qquad\qquad   \,\, p \in \mathbb{R}, \, 0< \alpha<2 \nonumber 
\end{eqnarray}
which defines a new class of shifted single particle operators and is a subject of interest for future research by its own.

\noindent
As a consequence we obtain the relation
\begin{eqnarray}
\label{eigen43}
\hat{C}^\dagger _{p}(k) &=& 
\frac{p^{-\alpha/2}\sin(\pi\alpha/4 ) \Gamma(1-\alpha/2)}{\sin(\pi\alpha/2 ) \Gamma(1-\alpha)}(
\hat{A}^\dagger_{p}(k) + \hat{B}^\dagger_{p}(k) + R(k,p)) \\
\hat{C}_{p}(k)  &=& 
\frac{p^{-\alpha/2}\sin(\pi\alpha/4 ) \Gamma(1-\alpha/2)}{\sin(\pi\alpha/2 ) \Gamma(1-\alpha)}(
\hat{A}_{p}(k) +\hat{B}_{p}(k)+ R(k,p)) \\
& & 
\qquad \qquad \qquad \qquad \qquad \qquad  \,\, p \in \mathbb{R},  \, \, 0< \alpha<1 \nonumber 
\end{eqnarray}
with a residual interaction term $R(k,p)$, which vanishes for $p \rightarrow \infty$. 

\noindent
With these definitions, we obtain for the Hamiltonian (\ref{EHHH}) :
\begin{eqnarray}
\label{EHHH2}
H &\approx& 
\frac{\sin^2(\pi\alpha/4 )  \Gamma^2(1-\alpha/2)}{\sin^2(\pi\alpha/2 ) \Gamma^2(1-\alpha)}
\times 
\nonumber \\
&&
\sum_{k,p>0} E_{k,p} \,  p^{-\alpha} \, 
(
 \hat{A}^\dagger_{p} \, \hat{A}_{p} +
 \hat{B}^\dagger_{p} \, \hat{B}_{p} +
G_{k,p} ( 
 \hat{A}^\dagger_{p} \, \hat{B}_{p} +
  \hat{B}^\dagger_{p} \, \hat{A}_{p}))
\end{eqnarray}
where the constant  factor $G_{k,p} \neq 1$ emulates  the influence of the residual interaction $R(k,p)$.

This is a special case of the general pairing Hamiltonian \cite{bar57},\cite{bel59},\cite{kit87}
\begin{equation}
\label{Epair}
H{_\textrm{\tiny{pair}}} = \sum_{k} E_{k} \hat{a}^\dagger_{k} \, \hat{a}_{k}
 + \omega_k  \hat{b}^\dagger_{k} \, \hat{b}_{k} +
 g_k(
  \hat{a}^\dagger_{k} \, \hat{b}_{k} +
  \hat{b}^\dagger_{k} \, \hat{a}_{k})
\end{equation}
which by performing a Bogoliubov transformation \cite{bog49} may be exactly diagonalized  into \cite{gre96}
\begin{equation}
\label{EBOg}
H^{'}_\textrm{\tiny{pair}}  = \sum_{k>0} \Omega_{k}  (\hat{\beta}^\dagger_{k} \, \hat{\beta}_{k}+\hat{\beta}^\dagger_{-k} \, \hat{\beta}_{-k})
\end{equation}
where the annihilation-operators $\hat{\beta}_k$ are given as 
\begin{eqnarray}
 \hat{\beta}_{k} &=& u_k  \hat{a}_{k} + v_k  \hat{b}^\dagger_{k} \\
 \hat{\beta}_{-k} &=& u_k  \hat{b}_{k} + v_k  \hat{a}^\dagger_{k} 
\end{eqnarray}
and the corresponding creation operators $\hat{\beta}^\dagger_k$:
\begin{eqnarray}
 \hat{\beta}^\dagger_{k} &=& u_k  \hat{a}^\dagger_{k} - v_k  \hat{b}_{k} \\
 \hat{\beta}^\dagger_{-k} &=& u_k  \hat{b}^\dagger_{k} + v_k  \hat{a}_{k} 
\end{eqnarray}
where $v^2_k, u^2_k$ with  $u^2 +v^2 = 1$ are the occupation probabilities for a particle/hole state and $\Omega_{k}$ is the quasi-particle energy of a particle-hole state.

We may therefore interpret objects, which are generated using symmetric Erd$\acute{\text{e}}$lyi-Kober type creation and annihilation-operators as compounds of Riesz type objects, which we call quasi-particles in analogy to the classical case of a pairing Hamiltonian. 

A fractional Cooper-pair of this type may then be characterized by a coherent momentum vector $k$, where its two components of Riesz type are determined by the two momenta $k\pm p/2$ coupled to $k$.  

In nuclear physics the difference in binding energies for even and odd nuclei is explained due to a residual interaction of pairing type for protons and neutrons respectively \cite{bel59}. In this case it is tempting to associate $k$ with
the Fermi-level $k_F$, where $p$ determines the momentum of a given particle-hole pair.  

Using symmetric Erd$\acute{\text{e}}$lyi-Kober type
integrals, a fundamental new aspect in treating pairing effects emerges within the concept of fractional calculus. The pairing effects are modeled via a Cassini- type kernel function, which allows  a new analytical approach.

\section{Conclusion}
We have derived the eigenfunctions and eigenvalues of a reflection symmetric Erd$\acute{\text{e}}$lyi-Kober type
integral. In terms of a non-local quasi-particle Hamiltonian we have investigated 
the properties of this operator and found strong analogies to the quasi-particle concept used to describe pairing effects in e.g. solid states and nuclear physics.

Symmetric Erd$\acute{\text{e}}$lyi-Kober type
integrals allow a direct physical interpretation in terms of a quasi-particle operators, where two momenta are coupled to $p=0$. 

Even more important, within the framework of fractional calculus, these operators allow a new approach for an alternative treatment of multi-particle quantum systems, where general pairing effects are modeled analytically using a Cassini-type kernel function.    

This opens a new promising area of future research  within the framework of generalized fractional calculus.
\section*{Acknowledgements}

We thank A. Friedrich and V.~S.  Kiryakova  for useful
discussions and suggestions.



 \smallskip

 \it

 \noindent
 GigaHedron\\
 Berliner Ring 80, D-63303 Dreieich, GERMANY
  \\[4pt]
e-mail: herrmann@gigahedron.com
\hfill Received: July 17, 2014 


\end{document}